\documentclass[prb,10pt,twocolumn,superscriptaddress]{revtex4-1}
\usepackage[dvipdfmx]{graphicx}
\usepackage{dcolumn}
\usepackage{bm}
\usepackage{color}
\usepackage{ulem}

\begin{document}
\newcommand{\bR}{\mbox{\boldmath $R$}}
\newcommand{\tr}[1]{\textcolor{red}{#1}}
\newcommand{\trs}[1]{\textcolor{red}{\sout{#1}}}
\newcommand{\tb}[1]{\textcolor{blue}{#1}}
\newcommand{\tbs}[1]{\textcolor{blue}{\sout{#1}}}
\newcommand{\Ha}{\mathcal{H}}
\newcommand{\mh}{\mathsf{h}}
\newcommand{\mA}{\mathsf{A}}
\newcommand{\mB}{\mathsf{B}}
\newcommand{\mC}{\mathsf{C}}
\newcommand{\mS}{\mathsf{S}}
\newcommand{\mU}{\mathsf{U}}
\newcommand{\mX}{\mathsf{X}}
\newcommand{\sP}{\mathcal{P}}
\newcommand{\sL}{\mathcal{L}}
\newcommand{\sO}{\mathcal{O}}
\newcommand{\la}{\langle}
\newcommand{\ra}{\rangle}
\newcommand{\ga}{\alpha}
\newcommand{\gb}{\beta}
\newcommand{\gc}{\gamma}
\newcommand{\gs}{\sigma}
\newcommand{\vk}{{\bm{k}}}
\newcommand{\vq}{{\bm{q}}}
\newcommand{\vR}{{\bm{R}}}
\newcommand{\vQ}{{\bm{Q}}}
\newcommand{\vga}{{\bm{\alpha}}}
\newcommand{\vgc}{{\bm{\gamma}}}
\newcommand{\Ns}{N_{\text{s}}}
\newcommand{\avrg}[1]{\left\langle #1 \right\rangle}
\newcommand{\eqsa}[1]{\begin{eqnarray} #1 \end{eqnarray}}
\newcommand{\eqwd}[1]{\begin{widetext}\begin{eqnarray} #1 \end{eqnarray}\end{widetext}}
\newcommand{\hatd}[2]{\hat{ #1 }^{\dagger}_{ #2 }}
\newcommand{\hatn}[2]{\hat{ #1 }^{\ }_{ #2 }}
\newcommand{\wdtd}[2]{\widetilde{ #1 }^{\dagger}_{ #2 }}
\newcommand{\wdtn}[2]{\widetilde{ #1 }^{\ }_{ #2 }}
\newcommand{\cond}[1]{\overline{ #1 }_{0}}
\newcommand{\conp}[2]{\overline{ #1 }_{0#2}}
\newcommand{\nn}{\nonumber\\}
\newcommand{\cdt}{$\cdot$}
\newcommand{\bra}[1]{\langle#1|}
\newcommand{\ket}[1]{|#1\rangle}
\newcommand{\braket}[2]{\langle #1 | #2 \rangle}
\newcommand{\bvec}[1]{\mbox{\boldmath$#1$}}
\newcommand{\blue}[1]{{#1}}
\newcommand{\bl}[1]{{#1}}
\newcommand{\red}[1]{\textcolor{black}{#1}}
\newcommand{\rr}[1]{{#1}}
\newcommand{\bu}[1]{\textcolor{black}{#1}}
\newcommand{\cyan}[1]{\textcolor{black}{#1}}
\newcommand{\fj}[1]{{#1}}
\newcommand{\green}[1]{{#1}}
\newcommand{\gr}[1]{\textcolor{green}{#1}}
\newcommand{\tgr}[1]{\textcolor{green1}{#1}}
\newcommand{\tgrs}[1]{\textcolor{green1}{\sout{#1}}}
\newcommand{\tmg}[1]{\textcolor{black}{#1}}
\newcommand{\tmgs}[1]{\textcolor{magenta}{\sout{#1}}}
\newcommand{\txb}[1]{\textcolor{blue}{#1}}
\definecolor{green}{rgb}{0,0.5,0.1}
\definecolor{green1}{rgb}{0,1.0,0.0}
\definecolor{blue}{rgb}{0,0,0.8}
\definecolor{cyan}{rgb}{0,0.8,0.9}
\definecolor{grey}{rgb}{0.3,0.3,0.3}
\definecolor{orange}{rgb}{1,0.5,0.25}
\preprint{APS/123-QED}

\title{
Modulated Helical Metals at
Magnetic Domain Walls of Pyrochlore Iridium Oxides
}
\author{Youhei Yamaji}
\email{yamaji@ap.t.u-tokyo.ac.jp}
\affiliation{Quantum-Phase Electronics Center (QPEC), The University of Tokyo, Hongo, Bunkyo-ku, Tokyo, 113-8656, Japan}
\affiliation{Department of Applied Physics, The University of Tokyo, Hongo, Bunkyo-ku, Tokyo, 113-8656, Japan}
\author{Masatoshi Imada}
\affiliation{Department of Applied Physics, The University of Tokyo, Hongo, Bunkyo-ku, Tokyo, 113-8656, Japan}
\date{\today}

\begin{abstract}
Spontaneous symmetry breakings, metal-insulator transitions,
and transport properties of magnetic-domain-wall states in pyrochlore iridium oxides
are studied by employing a symmetry adapted effective hamiltonian with a slab perpendicular
to the (111) direction of the pyrochlore lattice.
Emergent metallic domain wall, which
has  unconventional topological nature with a controllable and mobile metallic layer,
is shown to host Fermi surfaces
with modulated helical spin textures resembling Rashba metals.
The helical nature of the domain-wall Fermi surfaces is
experimentally detectable by anomalous Hall conductivity,
circular dichroism, and optical
Hall conductivity under external magnetic fields. 
Possible applications of the domain-wall metals
to spin-current generation and ``half-metallic" conduction
are also discussed.
\end{abstract}
\pacs{
}
\maketitle

\section{Introduction}
Emergent quantum phases
of pyrochlore iridium oxides $R_2$Ir$_2$O$_7$ ({\it R}: rare-earth elements)
have attracted broad interest~\cite{doi:10.1143/JPSJ.70.2880,doi:10.1143/JPSJ.80.094701,
PesinBalents,PhysRevB.83.205101,doi:10.1146/annurev-conmatphys-020911-125138}.
Previous theoretical studies have predicted
Weyl semimetals in non-collinear magnetic phases of Y$_2$Ir$_2$O$_7$~\cite{PhysRevB.83.205101},
and non-Fermi-liquid ground states~\cite{Abrikosov_beneslavskii,PhysRevLett.111.206401,PhysRevLett.113.106401,PhysRevX.4.041027} or
strong topological insulators as spontaneously symmetry breaking phases~\cite{PhysRevLett.100.156401,PhysRevLett.103.046811,
PhysRevB.79.245331,PhysRevB.82.075125,kurita2011topological}
in a paramagnetic pyrochlore iridium oxide Pr$_2$Ir$_2$O$_7$.

The non-collinear magnetic phase
called the all-in$-$all-out (AIAO) phase,
which was theoretically predicted for Y$_2$Ir$_2$O$_7$~\cite{PhysRevB.83.205101} and experimetally confirmed for
Nd$_2$Ir$_2$O$_7$~\cite{doi:10.1143/JPSJ.81.034709} and Eu$_2$Ir$_2$O$_7$~\cite{PhysRevB.87.100403},
shows other possible intriguing transport properties~\cite{PhysRevB.84.075129,PhysRevB.85.241101}. For example, thin 
films of the pyrochlore iridium oxides with the AIAO orders have been proposed to
exhibit
anomalous Hall effects~\cite{PhysRevLett.112.246402}.
The authors have predicted that
magnetic domain walls in the AIAO phase host metallic domain-wall states
characterized by a zero-dimensional class A Chern number~\cite{PhysRevX.4.021035}. 

Magnetic domain walls have been an ingredient of spintronics.~\cite{Parkin}
In both of old-fashioned magnetic bubble memories and cutting-edge magnetoresistive random  access memories,
magnetic domains themselves have conveyed information.
In contrast, 
while the domains in the AIAO orders of the pyrochlore iridium oxides
themselves remain functionless magnetic insulators,
the magnetic domain walls in the AIAO phases
have been predicted to offer tunable, controllable and mobile two-dimensional metallic layers~\cite{PhysRevX.4.021035}.

In the present paper,
we clarify the functions of the predicted
topological domain-wall metals
of the pyrochlore iridium oxides in more details.
First, spontaneous symmetry breakings and metal-to-insulator
transitions at the magnetic domain walls are examined.
A single (111) magnetic domain wall
in a simple tight binding model for the pyrchlore iridium oxide
with spin-orbit couplings is studied
by employing a slab geometry perpendicular to the (111) direction and an unrestricted Hartree-Fock approximation.
Next, the transport properties of the (111) domain wall are studied by using the Kubo formula.
Anomalous and optical Hall conductivities characterize the domain-wall metals.

We show that
the domain-wall metals are characterized by modulated helical two-dimensional Fermi surfaces.
Resembling Rashba-split Fermi surfaces~\cite{Rashba,PhysRevLett.5.371,Bychkov_Rashba} observed, for example, in
surface states of noble metals~\cite{PhysRevLett.77.3419},
inverted semiconductor heterostructures~\cite{PhysRevLett.78.1335},
and, recently, in a bulk semiconductor BiTeI~\cite{ishizaka2011giant},
spin/total-angular-momentum polarization tangential to Fermi surfaces
is realized in the modulated helical Fermi surfaces.
Differently from intensively studied ideal helical Fermi surfaces of topological insulator surfaces~\cite{xia2009observation,
hsieh2009tunable,PhysRevLett.103.266801},
which are \bu{described by a Dirac hamiltonian $v_{\rm F}[\vec{k}\times\vec{\hat{\sigma}}]_z$},
the spin/total angular moment of electrons are affected \cyan{by trigonal warpings and
have their components perpendicular to the two-dimensional plane}.

\cyan{We also show that there are two categories of domain wall metals, namely,
helical metals with and without degeneracy:}
\tmg{Degenerate helical metal is characterized by the two-fold degeneracy of the Fermi surface states at every momenta and the helical metal emerges by a spontaneous symmetry breaking of this degeneracy leading to the emergence of a magnetic moment at the domain wall and the splitting of the two-fold degenerate Fermi surface, which emerges as a first-order transition and is}
\red{evidenced by}
sudden changes in the anomalous Hall conductivities
of the domain-wall metals \tmg{and the jump in the density of states at the Fermi level}.
Splittings of the modulated helical Fermi surfaces \red{eventually} lead to a metal-to-insulator
transition when the ratio of the intra-atomic Coulomb repulsions and the kinetic energy
is \red{further increased}
by applying \red{negative} chemical or physical pressures.
The absolute value of the anomalous Hall conductivity itself, however,
is extremely sensitive to boundary conditions.

We also propose that the optical Hall conductivity
offers an experimental hallmark of the degenerate helical domain-wall metals.
In contrast to the optical Hall conductivities due to \tmg{the} formation of
Landau levels, the imaginary part of the optical Hall conductivity of
the degenerate helical metals show a continuum of optical Hall responses
with a finite width proportional to Zeeman energy and, hence, proportional to
external magnetic fields.

\textcolor{black}{
The organization of the present paper is the following.
In this paper, we focus on magnetic domain walls in a Hubbard-type effective hamiltonian of electrons
in $J_{\rm eff}$=1/2-manifold of iridium atoms that constitute a pyrochlore lattice.
The effective hamiltonian is introduced in Sec.~\ref{section:MM}.
With a slab geometry, electronic and transport properties of a single magnetic domain wall of $R_2$Ir$_2$O$_7$
are studied by employing the unrestricted Hartree-Fock approximation and the Kubo formula,
which are also described in Sec.~\ref{section:MM}.
The quantum phase transitions of the domain-wall metals are examined in Sec.~\ref{section:DWPD}.
Section \ref{section:DHM} and \ref{section:HM} are devoted to detailing the electronic and
transport properties of the domain-wall metals.
The weak topological nature proposed in Ref.~\onlinecite{PhysRevX.4.021035}
and possible application of the domain-wall metals are discussed in Sec.~\ref{section:D}.
Section \ref{section:S} is devoted to a summary of the present paper.}

\section{Model and Method}
\label{section:MM}
\subsection{Simple effective hamiltonian for electrons in $J_{\rm eff}$=1/2-manifold}

\begin{figure}[htb]
\centering
\includegraphics[width=8cm]{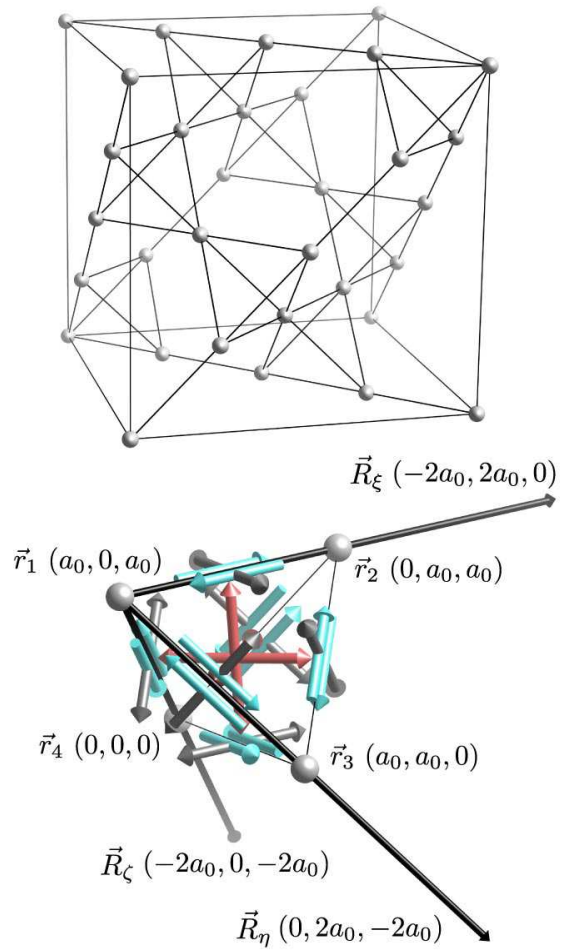}
\caption{(color online):
Cubic unit cell of a pyrochlore lattice (upper panel)
and illustration of $\vec{b}_{ij}$, $\vec{d}_{ij}$, and $\vec{b}_{ij}\times\vec{d}_{ij}$
in \tmg{Hamiltonian} $\hat{H}$ for a single tetrahedron consisting of four iridium atoms located at
$\vec{r}_{i}\ (i=1,2,3,4)$ (lower panel).
Arrows connecting $\vec{r}_j$ with $\vec{r}_i$ represent directions of $\vec{d}_{ij}$.
Vectors $\vec{b}_{ij}$ connect the center of the tetrahedron and the center of the bond.
Arrows perpendicular to $\vec{d}_{ij}$ show directions of $\vec{b}_{ij}\times\vec{d}_{ij}$.
Lattice vectors that define the slab geometry, $\vec{R}_{\xi}$, $\vec{R}_{\eta}$, and $\vec{R}_{\zeta}$,
are also shown.
}
\label{Figtetra}
\end{figure}

As \tmg{a simple} effective hamiltonian of pyrochlore iridium oxides, $R_2$Ir$_2$O$_7$,
we study the following effective Hubbard-type hamiltonian of electrons in
$J_{\rm eff}$=1/2-manifold of iridium atoms~\cite{PesinBalents,PhysRevB.82.085111,kurita2011topological,PhysRevX.4.021035}
on a pyrochlore lattice (see the upper panel of Fig.\ref{Figtetra}),
\eqsa{
\hat{H} &=&+t\sum_{i,j}^{\rm n.n.}\sum_{\sigma}
        \left[
        \hatd{c}{i\sigma}\hatn{c}{j\sigma}
        +{\rm h.c.}
        \right]
 +U\sum_{i}\hat{n}_{i\uparrow}\hat{n}_{i\downarrow}
        \nn
        &&-i\zeta \sum_{i,j}^{\rm n.n.}
          \sum_{\sigma,\sigma'=\uparrow,\downarrow}
          \hatd{c}{i\sigma}
          \left(\vec{\hat{\sigma}}\cdot
          \frac{\vec{b}_{ij}\times\vec{d}_{ij}}{|\vec{b}_{ij}\times\vec{d}_{ij}|}\right)_{\sigma\sigma'}
          \hatn{c}{j\sigma'},
\label{TB}
}
where a fermionic operator $\hatd{c}{i\sigma}$ ($\hatn{c}{i\sigma}$) creates (annihilates)
an electron
with $\sigma$-spin at $i$-th site.
Here, $\vec{\hat{\sigma}}=(\hat{\sigma}_x,\hat{\sigma}_y,\hat{\sigma}_z)^{T}$ is a vector consisting of
Pauli matrices,
$\vec{d}_{ij}$ is a vector from $j$-th site to $i$-th site,
and $\vec{b}_{ij}$ is a vector from the center of the tetrahedron including $i$-th and $j$-th sites
to the center of the bond connecting $i$-th and $j$-th sites (see the lower panel of Fig.\ref{Figtetra} that
illustrates
the directions of $\vec{b}_{ij}$, $\vec{d}_{ij}$, and  $\vec{b}_{ij}\times\vec{d}_{ij}$).
In the above hamiltonian, the spin-orbit couplings are decoded as
spin-dependent complex hoppings proportional to $\zeta$. 

The previous studies have shown that,
for $t>0$, $\zeta<0$, and $U/t \gg 1$,
the experimentally observed AIAO order is accounted for by the strong coupling expansion
of Eq.(\ref{TB}) at half filling: 
The second order terms proportional to $t\zeta/U$ lead to
the {\it \red{direct}} Dzyaloshinskii-Moriya (DM) interactions that
are known to stabilize AIAO orders~\cite{PhysRevB.78.214431,arXiv:1008.3038,PhysRevB.85.045124}.
The DM interactions determine not only the mutual angles
of the neighboring atoms' magnetic moments but also
the directions of these magnetic moments.
In the strong coupling limit, the magnetic moments
of the effective hamiltonian effectively feel Ising-like anisotropy
through the exchange coupling. 

Due to the effective Ising-like anisotropy,
quantum and thermal fluctuations
are suppressed and \tmg{the mean-field treatments are basically justified}
deeply inside the AIAO orders.
%
Thanks to the suppression,
the unrestricted Hartree-Fock approximation
becomes reasonable, which we employ
in this paper, as introduced below.

\subsection{Slab geometry}
To study magnetic domain walls of the pyrochlore iridium oxides $R_2$Ir$_2$O$_7$,
we introduce a slab with surfaces perpendicular to the (111) direction.
The slab consists of regularly stacked Kagom${\rm \acute{e}}$ and triangular layers.
Here, we assume \red{the} translational symmetry along the Kagom${\rm \acute{e}}$ and triangular layers
perpendicular to the (111) direction for simplicity.
Then, the slab is described as a two dimensional lattice with
a unit cell that contains iridium atoms at  
\eqsa{
\vec{R}_{a}=\vec{R}_{(n,\ell)}
=\vec{r}_{n}+(\ell -1)\vec{R}_{\zeta},
}
where $n$ $(n=1,2,3,4)$ is the site index inside a single tetrahedron,
$\ell$ $(\ell=1,2,\dots,L_{\rm th})$ is the layer index, $\vec{r}_n$ is defined with the distance
between nearest neighboring iridium atoms $\sqrt{2} a_0$ \cyan{(see also Fig.\ref{Figtetra})} as
\eqsa{
\vec{r}_1&=(a_0,0,a_0)^{T},\nn
\vec{r}_2&=(0,a_0,a_0)^{T},\nn
\vec{r}_3&=(a_0,a_0,0)^{T},\nn
\vec{r}_4&=(0,0,0)^{T},\nonumber
}
and
$\vec{R}_{\zeta}=(-2a_0,0,-2a_0)^{T}$ is a lattice vector of bulk pyroclore lattices, 
\tmg{which runs along the direction from $\vec{r}_1$ to $\vec{r}_4$}.
The two dimensional lattice is defined by the lattice vectors
$\vec{R}_{\xi}=(-2a_0,2a_0,0)^{T}$ and $\vec{R}_{\eta}=(0,2a_0,-2a_0)^{T}$.
Then, the top (bottom) surface of the slab is the Kagom${\rm \acute{e}}$ (triangular) layer (see Fig.\ref{Figtetra}).

We will show transport properties of the (111) magnetic domain walls later.
To define the transport coefficients of the domain-wall states,
here, we introduce an additional Cartesian coordinate
with the axes $X$, $Y$, and $Z$ are parallel to $(-2,1,1)/\sqrt{6}$, $(0,1,-1)/\sqrt{2}$, and $(1,1,1)/\sqrt{3}$, respectively.

\subsection{Unrestricted Hartree-Fock approximation}
Here, we employ
the unrestricted Hartree-Fock approximation,
to describe the AIAO phases and magnetic domain walls
by using
the effective hamiltonian $\hat{H}$ Eq.(\ref{TB}).

\cyan{We decouple the Hubbard interaction terms in $\hat{H}$ by introducing mean fields as,}
\eqsa{
 \hat{n}_{i\uparrow}\hat{n}_{i\downarrow}
 &\simeq&
 \left[\hatd{c}{i\uparrow},\hatd{c}{i\downarrow}\right]
 \left(
 \frac{\rho_i}{2}\hat{\sigma}_0
 -
 \frac{
 \vec{m}_i\cdot\vec{\hat{\sigma}}
 }{2}
 \right)
 \left[
 \begin{array}{c}
 \hatn{c}{i\uparrow}\\
 \hatn{c}{i\downarrow}\\
 \end{array}
 \right]
 \nn
 &&-
 \avrg{\hat{n}_{i\uparrow}}
 \avrg{\hat{n}_{i\downarrow}}
 +
 \avrg{\hatd{c}{i\uparrow}\hatn{c}{i\downarrow}}
 \avrg{\hatd{c}{i\downarrow}\hatn{c}{i\uparrow}},
}
where the mean fields are defined as
\eqsa{
 \rho_i &=& 
      \avrg{\hat{n}_{i\uparrow}}
      +
      \avrg{\hat{n}_{i\downarrow}},
 \\
 m_i^{x} &=& 
      \avrg{\hatd{c}{i\uparrow}\hatn{c}{i\downarrow}}
      +
      \avrg{\hatd{c}{i\downarrow}\hatn{c}{i\uparrow}},
 \\
 m_i^{y} &=& 
      -i\avrg{\hatd{c}{i\uparrow}\hatn{c}{i\downarrow}}
      +i\avrg{\hatd{c}{i\downarrow}\hatn{c}{i\uparrow}},
 \\
 m_i^{z} &=& 
      \avrg{\hat{n}_{i\uparrow}}
      -
      \avrg{\hat{n}_{i\downarrow}}.
}
We do not introduce any restrictions on these self-consistent mean fields $\rho_i$ and $\vec{m}_i$.

\subsection{Kubo formula}
\cyan{By employing the unrestricted Hartree-Fock approximation,
the hamiltonian $\hat{H}$ \tmg{is reduced} to the following general one-body hamiltonian,}
\eqsa{
\hat{H}_{\rm UHF}=
\sum_{\vec{k}}\sum_{a,b}h_{ab}(\vec{k})\hatd{c}{a\vec{k}}\hatn{c}{b\vec{k}},
}
\cyan{where lowercase Roman letters ($a,b,\dots$) specify the site and spin indices of electrons,
and $\vec{k}$ represents two dimensional lattice momenta perpendicular to the (111) direction.}
\cyan{In general, the one-body hamiltonian $\hat{H}_{\rm UHF}$ is diagonalized by
introducing an unitary matrix $U$ as,}
\eqsa{
\sum_{a,b}U_{a\alpha}^{\ast}(\vec{k})h_{ab}(\vec{k})U_{b\beta}(\vec{k})
=\delta_{\alpha,\beta}\epsilon_{\alpha},
}
\cyan{where lowercase Greek letters
such as $\alpha$ label
eigenstates of $\hat{H}_{\rm UHF}$
and $\epsilon_{\alpha}$ is
the corresponding
eigenvalue of $\hat{H}_{\rm UHF}$.}

\cyan{We calculate the transport properties of $\hat{H}_{\rm UHF}$ by using the Kubo formula.
In the present paper, we focus on charge conductivities $\sigma_{\mu\nu}(\omega)$ defiend below.
By using the
charge current operators}
\eqsa{
j_{\mu}^{\alpha\beta}(\vec{k})=
-\frac{e}{\hbar}\sum_{a,b}
U_{a\alpha}^{\ast}(\vec{k})
\frac{\partial h_{ab}(\vec{k})}{\partial k_{\mu}}
U_{b\beta}(\vec{k}),
}
\cyan{$\sigma_{\mu\nu}(\omega)$
\tmg{is given from the Kubo formula}~\cite{HuzioNakano,doi:10.1143/JPSJ.12.570} as}
\textcolor{black}{
\begin{widetext}
\begin{eqnarray}
\sigma_{\mu\nu} (\omega)
=
-i\frac{\hbar}{A}\sum_{\vec{k}}\sum_{\alpha,\gamma}^{\epsilon_{\alpha}\neq \epsilon_{\gamma}}
\frac{\left[f(\epsilon_{\alpha}(\vec{k})-E_{\rm F})-f(\epsilon_{\gamma}(\vec{k})-E_{\rm F})\right]j_{\mu}^{\gamma\alpha}(\vec{k})
j_{\nu}^{\alpha\gamma}(\vec{k})}
{(\epsilon_{\alpha}(\vec{k})-\epsilon_{\gamma}(\vec{k}))\left[\epsilon_{\alpha}(\vec{k})-\epsilon_{\gamma}(\vec{k})
+\hbar\omega+i\delta \right]}
+\frac{iD_{\mu\nu}}{\hbar\omega+i\delta},
\label{Kubof}
\end{eqnarray}
\end{widetext}
}
where the Drude part is given as
\eqsa{
D_{\mu\nu}=\frac{\hbar}{A}\sum_{\vec{k}}\sum_{\alpha}
\frac{\beta j_{\mu}^{\alpha\alpha}j_{\nu}^{\alpha\alpha}}{4\left\{\cosh \frac{\beta (\epsilon_\alpha (\vec{k}) -E_{\rm F} )}{2}\right\}^2},
}
and the area of the surface of the slab with $L^2$ unit cells is given as
$A=4\sqrt{3}a_{0}^{2} L^2$, $\beta$ is the inverse temperature, and $f(\epsilon)$ is the Fermi distribution function.
In this paper, we focus on zero temperature properties by taking $\beta$ sufficiently large.

\section{Domain-wall Phase Diagram}
\label{section:DWPD}
By employing the unrestricted Hartree-Fock approximation,
we calculate physical properties of (111) magnetic domain walls
of the effective hamiltonian Eq.(\ref{TB})
with a \red{typical} value of $\zeta/t$, $\zeta/t=-0.2$, and a slab geometry with 40 layers ($L_{\rm th}=40$).
Starting with an initial condition where
all-out (all-in) magnetic moments for the $\ell$-th tetrahedron for $1\leq \ell \leq 20$ ($21\leq \ell \leq 40$),
we obtain domain-wall solutions as local minima of \red{the} free energy.

\begin{figure}[hb]
\centering
\includegraphics[width=8cm]{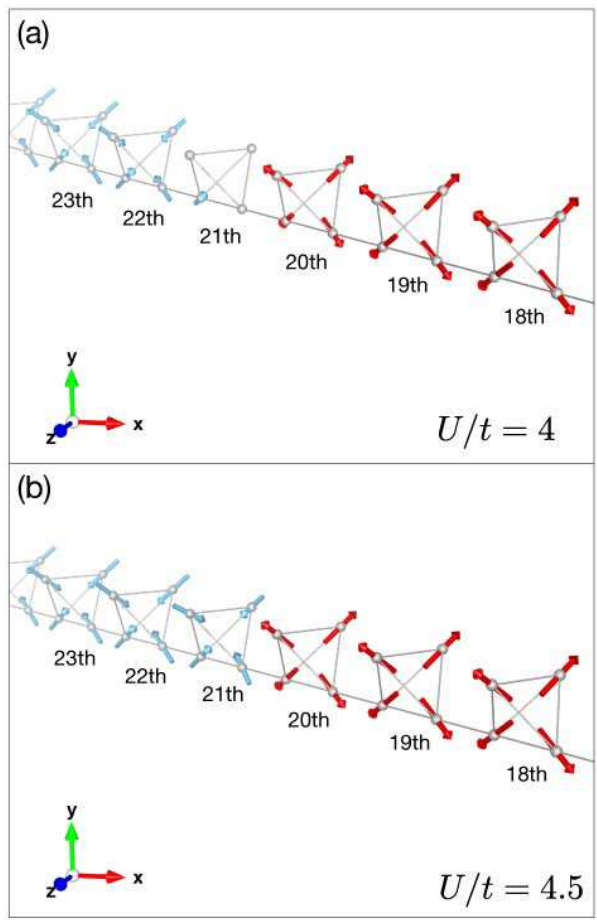}
\caption{(color online):
Magnetic moments of typical domain-wall solutions for (a) $U/t=4$
and (b) $U/t=4.5$.
For $U/t=4$, \tmg{the magnetic moment is zero} at
the Kagom${\rm \acute{e}}$-layer sites of the 21\tmg{st} layer.
}
\label{Figdw}
\end{figure}

As examples of the domain-wall solutions,
we illustrate the magnetic moments in Fig.\ref{Figdw} at $U/t=4$ and $U/t=4.5$.
At $U/t=4$, the domain-wall solution is symmetric under an inversion $I$ around
one of the Kagom${\rm \acute{e}}$-plane sites taken together with the time reversal operation $\Theta$, as is shown in an example of
Fig.\ref{Figdw}(a), where the 21\tmg{st} layer is the symmetric plane,
\textcolor{black}{if effects of surfaces of the slab with a finite thickness are neglected}.
\tmg{On the other hand,} this $I\Theta$-symmetry is spontaneously broken at $U/t=4.5$,
which is evident in evolution of \tmg{the} magnetization at the Kagom${\rm \acute{e}}$-plane sites of the 21\tmg{st} layer. Later, the magnetization at the 21\tmg{st} Kagom${\rm \acute{e}}$ plane \tmg{is denoted as the} domain-wall magnetization $m_{\rm dw}$
in comparison with the bulk magnetization $m_{\rm b}$ practically defined as the magnetization at the 10th or 30th layer.

\begin{figure}[ht]
\centering
\includegraphics[width=9cm]{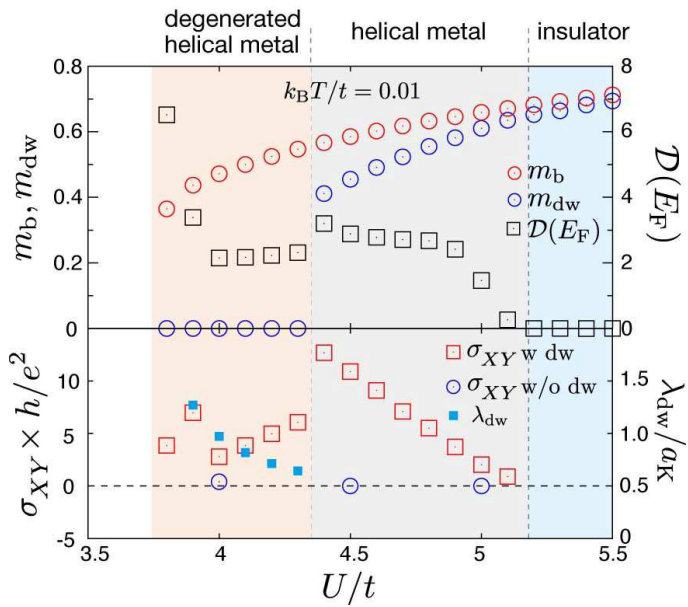}
\caption{(color online):
\tmg{Upper panel:} \tmg{$U/t$ dependence of} magnetization $(m_{\rm b},m_{\rm dw})$, and density of states $\mathcal{D}(E_{\rm F})$ \tmg{at the Fermi level}.
\tmg{Lower panel: $U/t$ dependence of} anomalous Hall
conductivity of a pyroclore slab with a single magnetic domain wall, and
domain-wall width $\lambda_{\rm dw}$ \red{in the unit of}
distances between nearest-neighbor
Kagom${\rm \acute{e}}$ layers $a_{\rm K}$. 
\cyan{In comparison with the anomalous Hall
conductivity with a single domain wall (shown as ``$\sigma_{\textcolor{black}{XY}}$ w dw"),
we also show the anomalous Hall
conductivity without domain walls shown as ``$\sigma_{\textcolor{black}{XY}}$ w/o dw."} 
\tmg{Three phases, degenerate helical metal, helical metal, and insulator are identified from the density of states.}
}
\label{Fig1}
\end{figure}

\red{There are actually three phases for the domain-wall states, a degenerate
helical metal, a helical metal, and an insulator.
The phase transition between the degenerate
helical and helical metals is of first order
accompanied by the spontaneous symmetry breaking of the $I\Theta$-symmetry \tmg{in the helical-metal phase,} 
\cyan{The metal-to-insulator transitions at the domain walls are \tmg{determined} by the \tmg{distinction between zero and nonzero} densities of states at the
Fermi level $E_{\rm F}$, $\mathcal{D}(E_{\rm F})$, which is given as}}
\eqsa{
\mathcal{D}(E_{\rm F})=
\frac{1}{L^2}\sum_{\vec{k}}\sum_{\alpha}
\frac{\beta}{4\left\{\cosh \frac{\beta (\epsilon_\alpha (\vec{k}) -E_{\rm F} )}{2}\right\}^2}.
}
\cyan{The $U/t$-dependences of $m_{\rm b}$, $m_{\rm dw}$, and $\cal{D}(E_{\rm F})$ are summarized in
the upper panel of Fig.~\ref{Fig1}.}

\cyan{In the lower panel of Fig.~\ref{Fig1}, we summarize anomalous Hall conductivities $\sigma_{\textcolor{black}{XY}}(\omega)$
of the two-dimensional domain-wall metals and domain-wall thickness $\lambda_{\rm dw}$
defined through fitting \tmg{the sublattice} magnetization
with $m_{\rm b}\tanh [(r-r_0)/\lambda_{\rm dw}]$, where $r-r_0$ is a distance from the 21\tmg{st Kagom${\rm \acute{e}}$ layer}
measured along the (111) direction.}
\cyan{The anomalous Hall conductivities for a slab with a single domain wall and without any domain walls
are shown in the lower panel of Fig.\ref{Fig1}.
In the $U/t$-dependence of $\sigma_{\textcolor{black}{XY}}$ with a single domain wall, the first order phase transition
between the degenerate helical metals and the helical metals is evident as a jump in $\sigma_{\textcolor{black}{XY}}$.
\tmg{The domain-wall thickness $\lambda_{\rm dw}$ can be fitted only in the degenerate helical metals with
the vanishing magnetization $m_{\rm dw}$, while the width of the domain wall is practically zero in the helical metal and insulator phases.}
For $U/t\lesssim 3.8$, the domain-wall structures
are not characterized by an assumption $m_{\rm b}\tanh [(r-r_0)/\lambda_{\rm dw}]$.}

\begin{figure*}[htb]
\centering
\includegraphics[width=16cm]{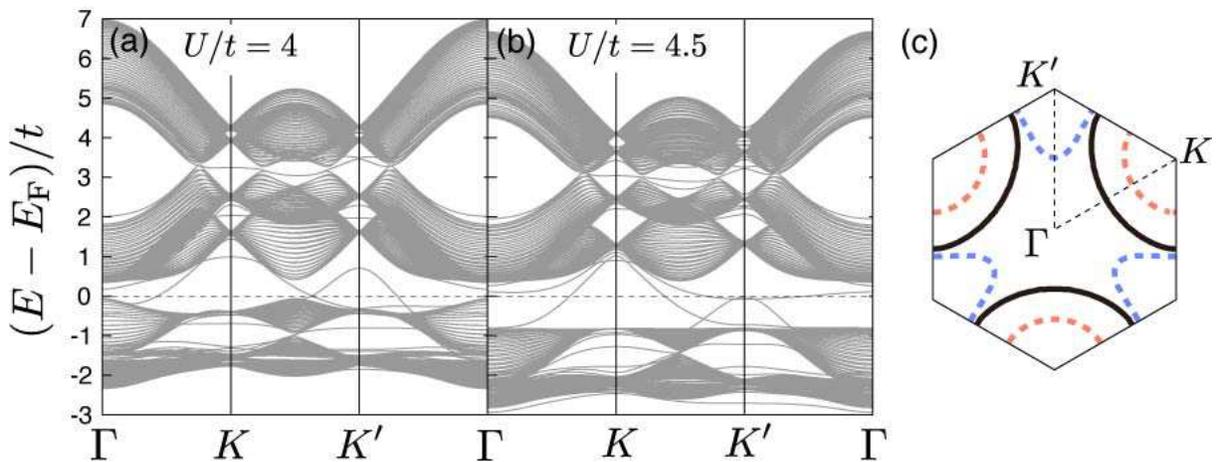}
\caption{(color online):
\cyan{Band dispersion of pyrochlore slab and Fermi surface. 
(a) and (b) show the
band dispersion of the slab for
$U/t=4$ and $U/t=4.5$, respectively.
The Fermi surfaces of \textcolor{black}{the domain-wall states} for $U/t=4$ and $U/t=4.5$ are
summarized in (c).
The solid curves represent the degenerate Fermi surfaces for $U/t=4$.
The broken (red and blue) curves represent \tmg{the split non-degenerate} hole pockets centered at the $K$ point
and electron pockets centered at the $K'$ point, \tmg{respectively} for $U/t=4.5$.} 
}
\label{FigFSbd}
\end{figure*}

\cyan{We next show
the band structures of the degenerate helical metals and the helical metals
in Fig.\ref{FigFSbd}. 
First of all, as naturally expected, there are no significant differences in
the \tmg{overall bulk} band dispersion for $U/t=4$ and $U/t=4.5$ as shown in Fig.\ref{FigFSbd}(a) and (b),
except for the bulk charge gaps, and the bands associated with the domain-wall and surface.
Ingap bands with band bottoms at the $K'$ point are surface bands of the slab.
The broken inversion symmetry (or the differences between the $K$ and $K'$ points)
of the slab in the AIAO phase is detectable in the domain-wall and surface states.
In contrast to the bulk band dispersion,
the spontaneous $I\Theta$-symmetry breakings at the domain wall
is evident in the domain-wall Fermi surfaces shown in Fig.\ref{FigFSbd}(c).} \tmg{Here, the $I\Theta$-symmetry breaking is evidenced from \textcolor{black}{the split Fermi surfaces centered at the $K$ and $K'$ pionts,
while these two Fermi surfaces are degenerated in the degenerate helical metals.}}

\begin{figure}[htb]
\centering
\includegraphics[width=4cm]{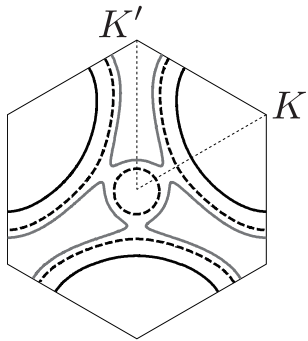}
\caption{(color online):
\cyan{Schematic illustration for
changes in Fermi surface topology.
Starting with a hole pocket centered at the $K$ point obtained in
the degenerate helical metals (solid black curves),
we illustrate how an electron pocket centered at the $K'$ point
emerges in the helical metals (solid gray curves).
Inbetween them, Fermi surfaces with two hole pockets
centered at the $K$ and $\Gamma$ points are illustrated.}
}
\label{FigLifshitz}
\end{figure}

\cyan{In the degenerate helical metals,
the Fermi surfaces consist of degenerated hole pockets centered at the $K$ point.
On the other hand, in the helical metals,
the Fermi surfaces consist of a hole pocket centered at the $K$ point and
an electron pocket centered at the $K'$ point.
The impact of the broken $I\Theta$-symmetry is significant
in the electron pocket centered at the $K'$ point.
\tmg{In the evolution} from the degenerated hole pocket centered at the $K$ point
to the electron pocket centered at the $K'$ point \tmg{with increasing $U/t$},
there should be changes in topology of the Fermi surface, as illustrated in Fig.\ref{FigLifshitz}.
Starting with the hole pockets centered at the $K$ point obtained in
the degenerate helical metals,
by increasing the hole density for the band that forms one of the hole pockets,
a hole pocket centered at the $\Gamma$ point \tmg{is also created}.
For further increase in the hole density, the newly-created hole pocket around the $\Gamma$ point
and the original hole pocket around the $K$ point touch each other.
Then, a closed electron-like Fermi surface enclosing the $K'$ point appears
\tmg{through a Lifshitz transition}\cite{Lifshitz60}.
Due to the
changes in the Fermi surface topology, \tmg{namely the Lifshitz transition},
the $I\Theta$-symmetry breakings become of first order~\cite{doi:10.1143/JPSJ.75.094719}.
}

\textcolor{black}{Before going into distinct natures of degenerate helical and helical metals,
which will be discussed in the following sections,
we show that the helical natures of these metallic states are evident in absorption of cicularly polarized light:
Circular dichroism
appears
in absorption of cicularly polarized light incident along the (111) direction on the domain walls,
in both of degenerate-helical- and helical-metal phases.
If we choose $t\sim 0.1$eV by comparing the band width of the present hamiltonian
with LDA results for Y$_2$Ir$_2$O$_7$~\cite{PhysRevB.83.205101},
we predict experimentally detectable circular dichroism in mid-wavelength or far infrared region.}

\textcolor{black}{In two-dimensional metals, the plasma frequency at small momentum $q$
is proportional to $|q|^{-1/2}$.
The simple Drude model for optical properties of single-band two-dimensional metals~\cite{TheisSS}, therefore,
predicts optical transparency. Even for the three dimensional systems,
if the thickness of the metallic layers is negligible, the intraband optical response becomes vanishing.
Consequently, the optical absorption of electromagnetic waves at the magnetic domain walls
is expected to be governed by and proportional to
the interband components of optical conductivities~\cite{doi:10.1143/JPSJ.20.412}.
Here, we examine the circular dichroism through {\it non-Drude part of}
optical conductivities for the circularly polarized light $\widetilde{\sigma}_{\pm\mp}(\omega)$. 
To obtain $\sigma_{\pm\mp}(\omega)$, we replace
$j^{\gamma\alpha}_{\mu}(\vec{k})j^{\alpha\gamma}_{\nu}(\vec{k})$
with $j^{\gamma\alpha}_{\pm}(\vec{k})j^{\alpha\gamma}_{\mp}(\vec{k})$
and set $D_{\mu\nu}=0$ in Eq.(\ref{Kubof}),
where $j^{\gamma\alpha}_{\pm}(\vec{k})=2^{-1/2}\left[j^{\gamma\alpha}_{\textcolor{black}{X}}(\vec{k})
\mp i j^{\gamma\alpha}_{\textcolor{black}{Y}}(\vec{k})\right]$.}

\begin{figure}[htb]
\centering
\includegraphics[width=8cm]{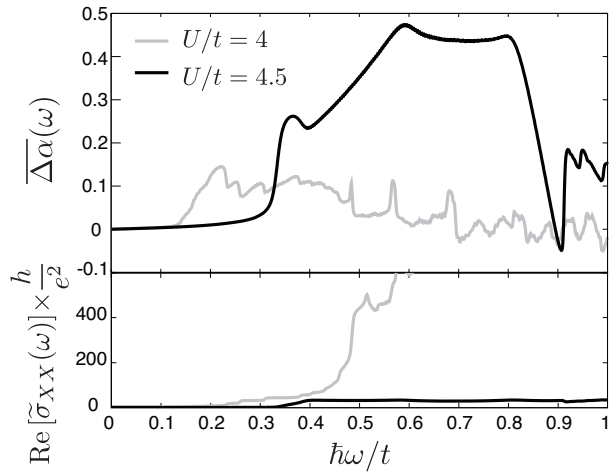}
\caption{(color online):
\textcolor{black}{
Circular dichroism in optical conductivity in comparison with
interband components of optical conductivity.
Upper panel shows frequency dependences of $\overline{\Delta\alpha}(\omega)$
for $U/t=4$ (grey solid curve) and $U/t=4.5$ (black solid curve).
Non-Drude parts or interband components of optical conductivity, $\widetilde{\sigma}_{XX}(\omega)$,
are shown in the lower panel.}
}
\label{FigCD}
\end{figure}

\textcolor{black}{
As an index to measure the circular dichroism, we choose a relative difference
between $\widetilde{\sigma}_{+-}(\omega)$ and $\widetilde{\sigma}_{-+}(\omega)$, defined as
\eqsa{
\overline{\Delta \alpha}(\omega)=
\frac{{\rm Re}[\widetilde{\sigma}_{+-}(\omega)]-{\rm Re}[\widetilde{\sigma}_{-+}(\omega)]}
{{\rm Re}[\widetilde{\sigma}_{+-}(\omega)]+{\rm Re}[\widetilde{\sigma}_{-+}(\omega)]}.
}
It shows substantial amplitudes of $\overline{\Delta \alpha}(\omega)$ around the optical absorption edge,
as we see in Fig.\ref{FigCD} (upper panel)
in comparison with interband components of $\widetilde{\sigma}_{XX}(\omega)$ (lower panel).
}

\section{Degenerate Helical Metals}
\label{section:DHM}
\cyan{In this section, we further clarify the nature of the degenerate helical
metals. First, we examine the spin polarization on the degenerated Fermi surfaces.
Next, we propose that
the \tmg{existence of the} degenerate helical metals can be \tmg{evidenced} in experiments by their characteristic frequency and magnetic-field dependences of the optical Hall conductivities. }
\subsection{Spin polarization on Fermi surfaces}
In this subsection, we show the
$Z_2$ nature of the spin alignments on the Fermi surface of the domain wall states.
First, we note that the degenracy of the eigenstate of $\hat{H}_{\rm UHF}$
is determined by the $Z_2$ symmetry for the spin on the Fermi surface, instead of the SU(2) symmetry seen in Fermi surfaces of usual paramagnetic metals.

\begin{figure}[ht]
\centering
\includegraphics[width=9cm]{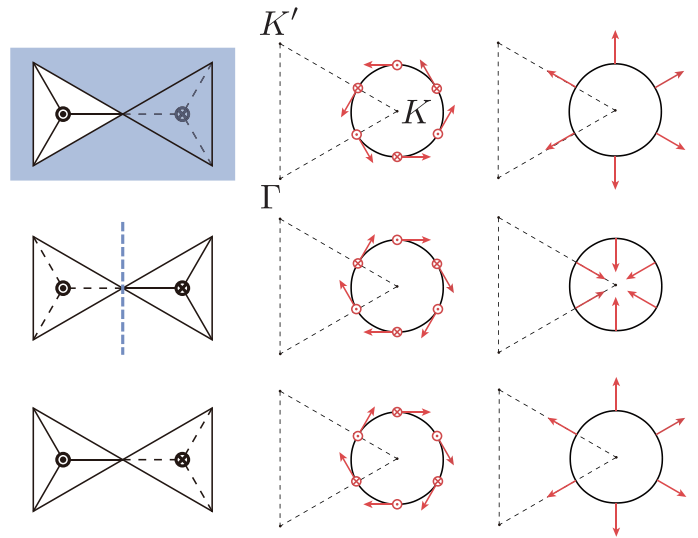}
\caption{(color online):
\cyan{Candidates of spin polarization imposed by symmetry operations on domain-wall Fermi surfaces.
In the left column, two tetrahedra on the Kagom${\rm \acute{e}}$ plane and the mirror planes
(shaded area and vertical broken line) are shown, where magnetic moments at the vertices are
illustrated as $\odot$ and $\otimes$.
Starting with spin polarization tangential and perpendicular to the Fermi surface that are shown in
the top panel of the middle and right column, respectively,
we operate the mirror reflections.
In the middle panel, we show the resultant spin polarization after the reflection with respect to
the Kagom${\rm \acute{e}}$ plane.
In the bottom panel, the resultant spin polarization after the two successive mirror reflections
is shown.} 
}
\label{FigS}
\end{figure}

This is intuitively understood from the simultaneously satisfied $I\Theta$ symmetry 
and the symmetry derived from a set of succesive two mirror operations, both of which are preserved around the domain wall as is easily seen from the magnetic structure of the AIAO domain wall as detailed below by using the illustrations in Fig.\ref{FigS}.

In the top panels of Fig.\ref{FigS}, we decompose the spin into the out-of-plane
(namely, along (111) direction) and the in-plane components by referring to the domain-wall plane,
where the in-plane component is further decomposed into the component tangential to the Fermi surface (more precisely the Fermi line)
and that perpendicular to the Fermi line in the two-dimensional Brillouin zone.

Let us first examine the degeneracy derived from the $I\Theta$ symmetry:
We first show that the  $I\Theta$ symmetry itself does not sufficiently restrict the direction of the spin on the Fermi surface.
The $I\Theta$ operation simply requires that, by this operation, all the spin components are transformed to the opposite directions.
The constraint is that the degenerate two states must have this spin inversion property.

However, the domain wall of the AIAO phase preserves another symmetry determined by the successive two mirror  operations illustrated step by step in the left three panels of Fig.~\ref{FigS} from the top to the bottom.
By this operation, the out-of-plane and tangential components of the spin are transformed to the opposite direction at least at the points where the Fermi surface (Fermi line) crosses
the symmetry lines denoted by the $\Gamma$-$K$, $K$-$K'$, and $\Gamma$-$K'$ lines. This inversion is the same as the $I\Theta$ operation
(see the bottom middle panel of Fig.~\ref{FigS}).
However, in contrast to the $I\Theta$ operation,
the component perpendicular to the Fermi surface within the domain-wall plane is unchanged under this operation as is seen in the  
bottom right panel of Fig.\ref{FigS}.
Therefore, the two symmetries are compatible only when the in-plane component perpendicular to the Fermi surface vanishes. This imposes a constraint that the in-plane component of the spin has only the tangential component at least in the illustrated symmetry points in the Brillouin zone. 

Since the spin direction is inverted by the $I\Theta$ and the two-mirror-plane operations, the spin degeneracy follows the $Z_2$ symmetry at each point of the Fermi surface.  At the different points on the Fermi surface, the continuity of the spin direction is necessary along the Fermi surface to lower the energy. The spin-orbit interaction indeed preserves the two-fold degeneracy of clock-wise and anti-clock wise direction of the helicity shown in Fig.\ref{FigS}, and the total $Z_2$ symmetry emerges.

\textcolor{black}{In other words, a generalization of Kramers degeneracy originating from
the $I\Theta$ and two-mirror symmetries of the domain-wall states, instead of the time-reversal symmetry,
guarantees the two-fold degeneracy of the eigenvalues of the one-body hamiltonian $\hat{H}_{\rm UHF}$ at $L_{\rm th}\rightarrow + \infty$.
The invariance of $\hat{H}_{\rm UHF}$ under the operation of $I\Theta$ and the two succesive mirror operations requires that
they map an eigenstate onto
another eigenstate with the same eigenvalues.
By noting that every single-particle eigenstate is a linear combination of $J_{\rm eff}=1/2$ orbitals
and $J_{\rm eff}=1/2$ states are not invariant under the operation of $I\Theta$,
the eigenstates of $\hat{H}_{\rm UHF}$ are generically mapped onto another eigenstate different from the original states under
the operation of $I\Theta$.
}

\textcolor{black}{By employing a realization of a $I\Theta$ operation, $R(I\Theta)$, we can write down the generalized Kramers degeneracy
explicitly as,
\eqsa{
\sum_{b}[R(I\Theta)]_{ab}U_{b\alpha}(\vec{k})
=e^{+i\theta_a}U_{a\beta}(\vec{k}),
}
and 
\eqsa{
\epsilon_{\alpha}(\vec{k})=\epsilon_{\beta}(\vec{k}),
}
where $\theta_a$ is a U(1) gauge of the wave function.
Then, we formulate an assumption that the real-space wave functions of the domain-wall states
are asymmetric around the inversion center of $I$.
The assumption is simply formulated as
\eqsa{
|U_{a\alpha}(\vec{k})|^2
\neq
|U_{a\beta}(\vec{k})|^2.
\label{assump1}
}
}

The exclusion of the spin polarization perpendicular to the Fermi surface (Fermi line)
clearly excludes a possibility that the domain-wall metals are simple paramagnetic metals.
It is a natural consequence
since, due to the strong spin-orbit couplings, the original SU(2) symmetry
of the electrons' spin is absent in the present system.

\subsection{An experimental hallmark of degenerate helical metals}

\cyan{The tangential spin polarization
introduces opposite helicities for the two degenerated Fermi lines.
The helicity inspires us to examine optical responses to incident light with circular polarization.
Indeed, by calculating the optical Hall responses, we found that
the low frequency behaviors of the optical Hall conductivities under external magnetic fields are
different from those of paramagnetic metals.}

\begin{figure}[hbt]
\centering
\includegraphics[width=9cm]{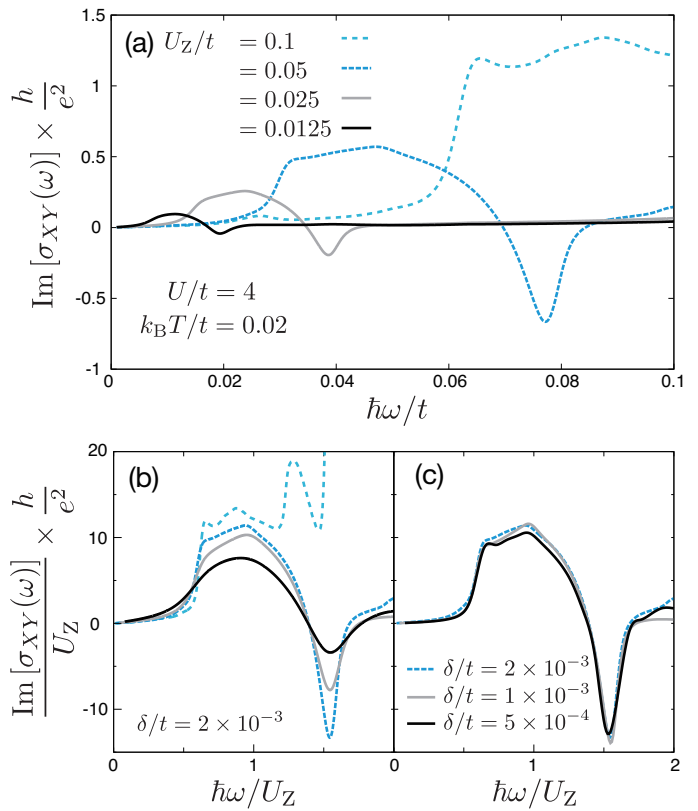}
\caption{(color online):
\red{(a)} Imaginary part of optical Hall conductivity of a pyrochlore slab with
a magnetic domain wall under several choices of external magnetic fields.
The optical Hall responses show continuum with finite width proportional to
Zeeman energy. 
\red{(b) Width of continuum shows proportionality to magnetic fields.
(c) Continuum indicating that introduced damping $\delta$ does not influence its structure.}
}
\label{Fig3}
\end{figure}

\red{The calculated imaginary part of the optical Hall conductivities are shown in Fig.\ref{Fig3} \cyan{for $U/t=4$}.
\cyan{In Fig.\ref{Fig3}(a), we show the low frequency behaviors of ${\rm Im}\sigma_{XY}(\omega)$ under external magnetic
fields $B$ parallel to the (111) direction.
The effect of the magnetic fields is taken into account only through the Zeeman term, because,
as we discuss later the effect from coupling through the electronic orbital motion is one order of magnitude smaller than the Zeeman effect, if we assume the electron effective mass in the order of $10 m_e$ with $m_e$ being the bare electronic mass,
as is anticipated from the bandwidth of the LDA calculation\cite{PhysRevB.83.205101,doi:10.7566/JPSJ.84.073703}.
By
changing the Zeeman \textcolor{black}{term,
\eqsa{
U_{\rm Z}
\sum_{i}\sum_{\sigma,\sigma'=\uparrow,\downarrow}
\hatd{c}{i\sigma}
\left[\textcolor{black}{\frac{\hat{\sigma}_x+\hat{\sigma}_y+\hat{\sigma}_z}{\sqrt{3}}}\right]_{\sigma\sigma'}
\hatn{c}{i\sigma'}
}}
where \textcolor{black}{$U_{\rm Z}$ is the Zeeman energy,}
the absorption continuum in the optical Hall conductivity $\sigma_{\textcolor{black}{XY}}(\omega)$
shifts its location
and width in frequency.
\textcolor{black}{Here, the Zeeman energy $U_{\rm Z}$ for $R_2$Ir$_2$O$_7$ is
estimated as
\eqsa{
U_{\rm Z}=
m_J g \mu_{\rm B} B + 6m_J \mathcal{J}_{R{\rm \mathchar`- Ir}}\mu_{\rm B}^{-1} m_{R}(B),
}
where $m_J$ ($g$) is a total magnetic moment ($g$-factor)
of an electron in $J_{\rm eff}=1/2$-manifold, $\mu_{\rm B}$ is the Bohr magneton,
$\mathcal{J}_{R{\rm \mathchar`- Ir}}$ is an exchange coupling between 
an iridium atom and the 6 neighboring $R$ atoms,
and $m_{R}(B)$ is the magnetic moment of the $R$ atoms
under external magnetic fields $B$.}
We note that the simplified isotropic exchange coupling $\mathcal{J}_{R{\rm \mathchar`- Ir}}$ is
employed, instead of the exchange couplings based on the symmetries of the wave functions of $R^{3+}$ ions~\cite{PhysRevB.86.235129,PhysRevB.89.075128}.
By rescaling the imaginary part of $\sigma_{XY}(\omega)$ and $\hbar\omega$ with the Zeeman energy
\textcolor{black}{$U_{\rm Z}$,}
in Fig.\ref{Fig3}(b), we show that the width of the continuum is proportional to the Zeeman energy
\textcolor{black}{$U_{\rm Z}$.}
Due to the finite line width introduced in Fig.\ref{Fig3}(a) and (b), $\delta/t=2\times 10^{-3}$, the detailed structures
of the continua are smeared.
To clarify the intrinsic structures of the continua, we change $\delta$ for each
value of the Zeeman energy with keeping the ratio of the Zeeman energy
\textcolor{black}{$U_{\rm Z}$}
and $\delta$,
as shown in Fig.\ref{Fig3}(c).
For three different values for the Zeeman energy
\textcolor{black}{$U_{\rm Z}/t=0.05, 0.025, 0.0125$},
the low frequency parts of ${\rm Im}\sigma_{XY}(\omega)$ are nearly collapsed into a single curve.
The collapse indicates that the amplitudes of ${\rm Im}\sigma_{XY}(\omega)$ are also
scaled with the Zeeman energy \textcolor{black}{$U_{\rm Z}/t$}. 
}}

\cyan{In usual metals, the imaginary part of the optical Hall conductivity arises
due to the formation of the Landau levels.
When the Landau levels induce the absorption in the optical Hall responses,
the absorption spectra consist of sharp peak structures, instead of forming continua.
In addition, under stronger external magnetic fields, the peak width is expected to remain unchanged
or to become narrower.
It is in sharp contrast with the present results.}

\textcolor{black}{Aside from the absorption peak shapes,
the energy scale of the absorption spectra in the degenerate helical metals is distinct from
that of usual paramagnetic metals.
While the energy scale of the absorption continuum is governed by the Zeeman energy $U_{\rm Z}$,
optical transitions between the Landau levels are governed by the cyclotron frequency,
\eqsa{
\hbar\omega_{\rm c}=\frac{2\mu_{\rm B}B}{m_{\rm eff}/m_{\rm e}},
}
where $m_{\rm eff}$ is the effective mass of the domain-wall states and
$m_{\rm e}$ is the electron mass.
In the present domain-wall states, we roughly estimate $m_{\rm eff}/m_{\rm e} \gtrsim \mathcal{O}(10)$.
Therefore, we expect that there are two distinct energy scales in optical Hall spectra,
namely, the Zeeman energy $U_{\rm Z}$ and the cyclotron mass $\hbar\omega_{\rm c}$,
which is roughly one order of magnitude smaller than $U_{\rm Z}$.
The expected separation of these two energy scales enables us
to distinguish helical metals and ordinary paramagnetic metals.
}

\cyan{For experimental observation, we roughly examine the energy scales
of the frequency and the required magnetic fields.
Here, by comparing the band width with pyrochlore iridium oxides and that of
the effective hamiltonian, we estimate the energy scale as $t\sim 0.1$ eV.
Therefore, the relevant frequency \tmg{range for the experimental observation revealed in Fig.~\ref{Fig3}(a)}  is up to 10 meV and
the amplitude of external magnetic fields is up to $10^2$ T.
The requirement for the external magnetic fields is seemingly demanding.
However, in the pyrochlore iridium oxide with $R=$Nd, for example,
the Zeeman fields originating from the magnetic moment of Nd atoms, \textcolor{black}{$m_{R={\rm Nd}}$},
assist the external magnetic fields in \tmg{enhancing} the Zeeman splitting of the degenerated
helical Fermi lines.
Thus, we expect that external magnetic fields up to several T are enough
\tmg{for detecting} the \tmg{characteristic} optical Hall responses discussed above in Nd$_2$Ir$_2$O$_7$.}

\begin{figure*}[ht]
\centering
\includegraphics[width=13cm]{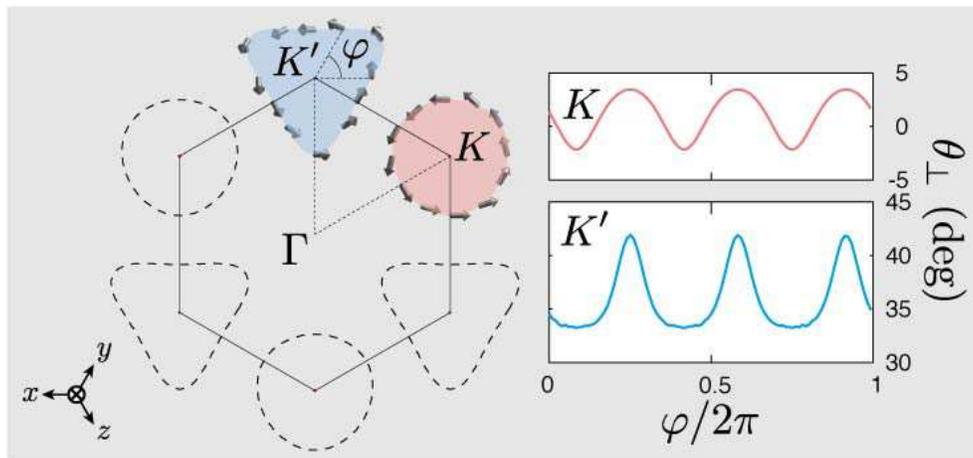}
\caption{(color online):
Spin polarization on electron and hole Fermi surfaces
around $K'$ and $K$ points, respectively,
determined by
spin-resolved spectral weight
$\vec{\mathcal{A}}\equiv
(\mathcal{A}_{x}(\vec{k},\omega),\mathcal{A}_{y}(\vec{k},\omega),\mathcal{A}_{z}(\vec{k},\omega))$
at $\omega=E_{\rm F}$ for
$U/t=4.5$ \red{in helical metal phase}.
The
spin polarization is illustrated as
arrows.
The out-of-plane components $\mathcal{A}_{111}\equiv
(\mathcal{A}_{x}+\mathcal{A}_{y}+\mathcal{A}_{z})/\sqrt{3}$
\cyan{are characterized by}
$\theta_{\perp}=\arcsin \left(
\mathcal{A}_{111}/|\vec{\mathcal{A}}|
\right)$,
in the right panel.
The magnetic domain wall keeps the three-fold symmetry around
(111) axis within entire range of $U/t$ in Fig.~\ref{Fig1}.
\red{The horizontal axis ($X$-axis) \tmg{in the left panel} is parallel to $(-2,1,1)/\sqrt{6}$, and
the vertical axis ($Y$-axis) is parallel to $(0,1,-1)/\sqrt{2}$.
The hexagonal Brillouin zone corners are $(k_X,k_Y)=(0,2\pi/3a)$ and
$(k_X,k_Y)=(\pi/\sqrt{3}a,\pi/3a)$.}
}
\label{Fig2}
\end{figure*}

\section{Helical Metals}
\label{section:HM}
\cyan{The spontaneously broken $I\Theta$-symmetry induces
split Fermi lines with spin polarization tangential to the Fermi lines with
additional components parallel to the (111) direction.
The explicit patterns of the spin polarization on the Fermi surfaces
are given in Fig.\ref{Fig2} for $U/t=4.5$.
As we mentioned in Fig.\ref{FigFSbd}, a hole pocket around the $K$ point and the electron pocket around the $K'$ point emerge.}

\cyan{Here, to illustrate the spin polarization quantitatively, we introduce}
\cyan{spin-resolved spectral weights. For example, the $x$-component of the spin-resolved
spectral weight is given by}
\eqsa{
  \mathcal{A}_{x}(\vec{k},\omega)
  =
  -\frac{1}{\pi}{\rm Im} \sum_{i\in \Gamma_{\rm dw}}\sum_{\sigma,\sigma'}\sum_{\alpha}
  \frac{U^{\ast}_{i\sigma\alpha}(\vec{k})(\hat{\sigma}_{x})_{\sigma \sigma'}U^{\ }_{i\sigma'\alpha}(\vec{k})}
  {\hbar\omega+i\delta - \epsilon_{\alpha}(\vec{k})},
  \nn
}
where $\Gamma_{\rm dw}$ is a subsystem consisting of 19th, 20th, 21\tmg{st}, and 22\tmg{nd} layers including a single domain wall.

\cyan{The spin polarization patterns shown in Fig.\ref{Fig2} resemble the spin-polarized Fermi surfaces
of the Rashba-split Fermi surfaces\cite{Rashba,PhysRevLett.5.371,Bychkov_Rashba} observed, for example, in BiTeI~\cite{ishizaka2011giant}.
However, as evident in Fig.\ref{Fig2}, the hole band dispersion is not captured by
the Rashba hamiltonian.}

\section{Discussion}
\label{section:D}
\subsection{Weak topological nature}
One might suspect that the emergence of insulating domain-wall states is inconsistent with
the existence of the 1D hidden {\it weak} Chern number proposed in Ref.\onlinecite{PhysRevX.4.021035}.
\textcolor{black}{The hidden topological nature is based on symmetry of the eigenstates of the mean-field
hamiltonian $\hat{H}_{\rm UHF}$ with a single AIAO magnetic domain wall:
The eigenstates at the $\Gamma$ point are invariant or constituting two-dimensional irreducible representation
under $C_3$ rotational symmetry around the (111) axis.
The {\it eigenvalues} of the $C_3$ rotation classify these eigenstates at the $\Gamma$ point into three categories.
Then, the number of the eigenstates in each categories gives us zero-dimensional Chern number.
When the (111) domain walls are shifted by a unit layer along the (111) direction, the zero-dimensional Chern number
is required to change.
The changes in the Chern number require the existence of the ingap domain-wall states.
In addition to the hidden Chern number, the $I\Theta$ symmetry of the AIAO domain walls in the degenerated helical metal phases
guarantee the existence of the metallic domain-wall states.}
However, in the helical metal phase and beyond (see Fig\ref{Fig1}. for the phase diagram),
the $I\Theta$ symmetry is broken and the topological protection is not guaranteed any more.

Even in the insulator phase, the insulating nature is fragile:
if the domain walls are deformed and translated by a single layer with keeping $C_3$
rotational symmetry around the (111) axis,
the changes in the zero dimensional Chern number
require that the two domain-wall bands touch each other at the $\Gamma$ point,
at least, once during the deformation.
In other words, the mutual touching of these domain-wall bands changes
the zero dimensional Chern number.
This asserts that the insulating phase may have gapless metallic
conduction to some extent with ``bad insulating" nature when the domain walls
are not strictly \textcolor{black}{flat, for example}.

\subsection{Memory effects in Hall measurements}

\red{Insertion of magnetic domain walls affects Hall conductivity}\textcolor{black}{,
as evident in comparison between the Hall conductivity of a single domain slab and a slab with
two domains that is separated by a single domain wall (Fig.\ref{Fig1}).}
Therefore, hysteresis due to insertion and removal of magnetic domain walls
in $R_2$Ir$_2$O$_7$, which is observed in magnetoresistivity~\cite{doi:10.7566/JPSJ.82.023706,PhysRevB.87.060403,PhysRevB.89.075127},
inevitably accompanies hysteresis in Hall conductivity.

\subsection{Spin current generation by charge current injection}
\textcolor{black}{In the helical-metal phase of the domain wall,
there are a hole pocket centered at the $K$ point and an electron pocket centered at 
the $K'$ point with modulated helical spin polarizaton.
These two helical Fermi pockets stimulate interest in possible application to
spintronics devices.}

\textcolor{black}{As same as the Rashba metals, spin polarization of electronic
and hole carriers generated by charge current injection are anti-parallel
if the Fermi velocity of these carriers are along the same
direction, leading to the cancellation of the spin current
for the contribution from the same Fermi velocity.
However, due to lack of the $I$-symmetry at the domain walls,
the Fermi velocities of the hole carriers around the $K$ point and
electron carriers around the $K'$ point are, in general, different.
Consequently, the difference in the Fermi velocities of two type of carriers generates
spin currents under the external charge current injection.
In the present simple tight-binding hamiltonian,
the difference is significant, as seen in the band dispersion of the helical metals in Fig.\ref{FigFSbd}(b).}

\subsection{``Half-metals" from lightly-doped domain-wall insulators}
\begin{figure}[bht]
\centering
\includegraphics[width=7cm]{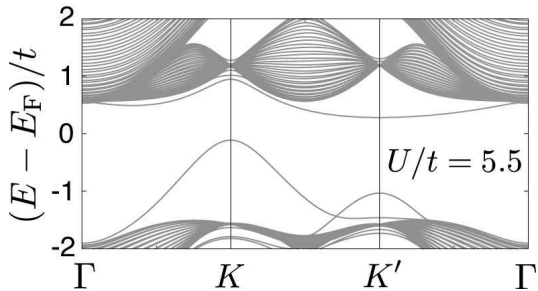}
\caption{(color online):
Band dispersion of insulating domain-wall states for $U/t=5.5$.
}
\label{dwbd55}
\end{figure}
\textcolor{black}{Finally, we discuss about a possible functionality of the domain-wall insulators
in $R_2$Ir$_2$O$_7$.
As briefly explained below, lightly doped domain-wall insulators are expected to have a property equivalent to half metals.}

\textcolor{black}{
In the insulating phases for $U/t\gtrsim 5.2$ (see the phase diagram shown in Fig.\ref{Fig1}),
the top (bottom) of the hole (electron) band is located at the $K$ ($K'$) point.
In Fig.\ref{dwbd55}, the slab band dispersion is shown for $U/t=5.5$ as an example of the insulating domain-wall states.
At the both symmetric points in the Brillouin zone of the domain-wall states,
the $C_3$ rotational symmetry around the (111) direction guarantees
the full spin polarizations of the Bloch state along the (111) direction, at the two $\vec{k}$ points:
The invariance of the $K$ and $K'$ under the $C_3$ rotation around the (111) axis prohibits the spin polarization
within the (111) plane that breaks the invariance under the $C_3$ rotation.
Therefore, if electronic or hole carriers are doped slightly into the domain walls,
the doped carriers in the domain-wall insulators become completely spin polarized in the (111) direction,
which is equivalent to electronic or hole half metals, respectively.}

\section{Summary}
\label{section:S}
Quantum phase transitions, electronic and transport
properties of domain-wall states of all-in$-$all-out
magnetic orders in pyrochlore lattice iridium oxides $R_2$Ir$_2$O$_7$ ($R$: rare-earth elements)
are studied by using a symmetry adapted Hubbard-type hamiltonian of $J_{\rm eff}$=1/2-manifold in the present paper.
There exist three quantum phases at the magnetic domain walls:
Metallic phases with degenerated helical Fermi surfaces, metals with helical electronic and hole Fermi pockets,
and insulators appear at the domain walls depending on the ratio of the intra-atomic Coulomb repulsion $U$ to
the nearest-neighbor hopping matrix $t$.
Gapless charge excitations in the degenerated helical metals are guaranteed by degeneracy
due to $I\Theta$-symmetry of the self-consistent domain-wall solutions
combined with the existence of a weak one-dimensional Chern number,
which is defined at the $\Gamma$ point of the domain-wall Brillouin zone.
Experimental hallmarks of the degenerated helical metals are also examined.
Spin polarization on the Fermi surfaces in the degenerated helical and helical metals
is detailed by employing the double group symmetry of the domain walls and
by showing direct numerical results.
Possible applications of these domain-wall states such as spin current generation
in the helical metals and half-metallic-like properties of lightly doped domain-wall insulators
are also discussed.

\acknowledgments
The authors thank Kentaro Ueda, Taka-hisa Arima, and Yoshinori Tokura
for fruitful discussion.
We acknowledge the financial supports
by a Grant-in-Aid for Scientific Research (Grants No. 22104010
and No. 22340090) from MEXT, Japan.
Y. Y acknowledge the financial supports
by a Grant-in-Aid for Young Scientists (B) (Grant No.
15K17702) from MEXT, Japan.
This work was also
supported by the Strategic Programs for Innovative Research (SPIRE), MEXT
conducted by the RIKEN Advanced Institute for Computational Science (AICS)
(Grants No. hp130007 and hp140215) and Computational Materials
Science Initiative (CMSI), Japan.
The lattice structure and magnetic moments were visualized using
VESTA 3~\cite{Momma}.

\bibliography{AHM2}
\end{document}